\documentclass[letter]{aa} 

\usepackage{graphicx}
\usepackage{txfonts}
\usepackage{wasysym}
\usepackage{ulem}
\begin{document}
\title{Glory revealed in disk-integrated photometry of Venus}

   \author{A. Garc\'ia Mu\~noz
          \inst{1}\fnmsep\thanks{Corresponding author: tonhingm@gmail.com}
          \and
          S. P\'erez-Hoyos\inst{2,3}
          \and
          A. S\'anchez-Lavega\inst{2,3}
          }

   \institute{ESA Fellow, ESA/RSSD, ESTEC, 2201 AZ Noordwijk, The Netherlands
    \and
    Grupo de Ciencias Planetarias, Departamento de F\'isica Aplicada I, 
    ETS Ingenier\'ia, UPV/EHU, Bilbao, Spain    
    \and
    Unidad Asociada Grupo de Ciencias Planetarias, UPV/EHU-IAA/CSIC, Spain
    }



  \abstract 
  {
  Reflected light from a spatially unresolved planet yields
  unique insight into the overall optical properties of the planet cover. 
  Glories are optical phenomena caused by light that is backscattered within spherical
  droplets following a narrow distribution of sizes;
  they are well known on Earth as localised features above liquid clouds. 
  } 
  {
  Here we report the first evidence for a glory in the disk-integrated 
  photometry of Venus and, in turn, of any planet.
  } 
  {We used previously published phase curves of the planet 
  that were reproduced over the full range of phase angles
  with model predictions based on a realistic description of the Venus atmosphere.  
  We assumed that the optical properties of the planet as a whole
  can be described by a uniform and stable cloud cover, an assumption that agrees well
  with observational evidence.
  } 
  {  We specifically show that the measured phase curves 
  mimic the scattering properties of the Venus 
  upper-cloud micron-sized  aerosols, 
  also at the small phase angles at which the glory occurs,  
  and that the glory contrast is consistent 
  with what is expected after multiple scattering of photons. 
  In the optical, the planet 
  appears to be brighter at phase angles of $\sim$11--13$^{\circ}$ than
  at full illumination; it undergoes a maximum dimming of up to 
  $\sim$10\%  at phases in between. 
  }
  {
  Glories might potentially indicate spherical droplets and, thus, extant 
  liquid clouds in the atmospheres of exoplanets. 
  A prospective detection will require exquisite photometry 
  at the small planet-star separations of the glory phase angles. 
  }


   \titlerunning{Glory in the disk-integrated phase curves of Venus}
   \maketitle
%

\section{Introduction}

Clouds occur regularly in the atmospheres of the solar system planets 
but it remains extremely challenging to predict where and how they 
form and what the fundamental properties of the resulting cloud particles
will be (Rossow et al., \cite{rossow1978}; S\'anchez-Lavega et al., 
\cite{sanchezlavegaetal2004}). 
Clouds influence the overall energy balance of a planet, as well as the
atmospheric dynamics and surface temperature.  
Clouds also have an effect on the planet's appearance as viewed from afar
and on our capacity to identify the atmospheric gases. 
Finding tools for the remote characterisation of clouds 
has long been a key aspect of planetary research
(Coffeen et al., \cite{coffeen1979}; Knollenberg et al., \cite{knollenbergetal1977}).

In that respect, Venus is quite unique because it is fully and
permanently covered by a complex cloud system (Esposito et al., \cite{espositoetal1983})   
and because it can be investigated from Earth over the entire range of 
Sun-Venus-observer phase angles, 
also in measurements of the planet's disk-integrated phase curve.
Unlike polarisation phase curves, which can be rich in diagnostic features 
that provide information about the fundamental properties of the scattering particles in the
atmosphere (Bailey, \cite{bailey2007}; 
Hansen \& Arking, \cite{hansenarking1971}; 
Hansen \& Hovenier, \cite{hansenhovenier1974};
Zugger et al., \cite{zuggeretal2010, zuggeretal2011}),
photometry is often less informative (Arking \& Potter, \cite{arkingpotter1968}). 
Features that appear distinctly in the polarisation phase curves of Venus
such as the glory, primary rainbow, or anomalous
diffraction (Hansen \& Arking, \cite{hansenarking1971}; Hansen \& Hovenier, \cite{hansenhovenier1974}) 
were key in the investigation of the Venus clouds.
Polarimetry is technically more challenging than photometry, however, 
which is the reason why the latter has traditionally been favoured  in studies of planetary atmospheres. 
Interestingly, the emerging research field of exoplanets is leading to a re-evaluation of
the polarimetry potentialities because it offers unique advantages in separating the 
planet from the (largely unpolarised) stellar glare.
Future efforts may need to combine both techniques to overcome the inherent difficulties
in this new field. 
We here investigate the photometric phase curves of Venus that reveal distinct evidence for a glory, 
which is an optical phenomenon associated with backscattering within spherical droplets.

\section{Venus phase curves}

\begin{figure*}
   \centering
   \includegraphics[width=15cm]{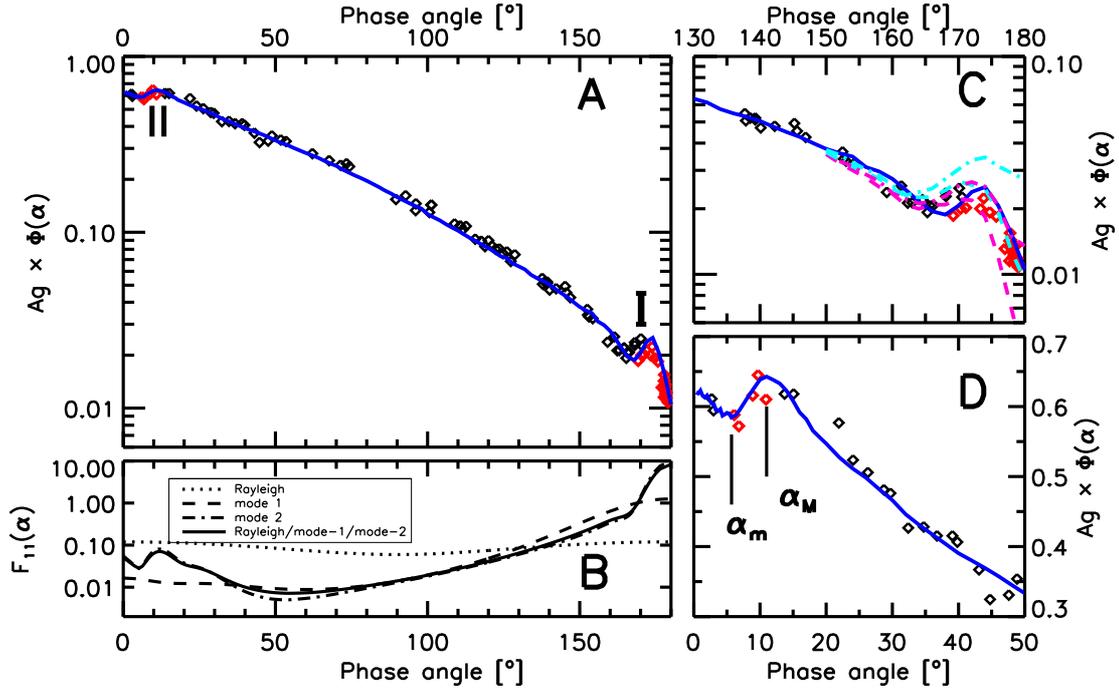}
      \caption{\label{LCmallama_fig}
      Venus phase curve in the B band and scattering particle phase functions 
      at $\lambda$=0.445 $\mu$m. 
      \textbf{A}. Symbols stand for the measured phase curve. 
      Black and red are ground- and space measurements. 
      The solid blue line is our modelled phase curve at $\lambda$=0.445 $\mu$m. 
      \textbf{B}. Scattering phase functions for Rayleigh particles, 
      mode-1 and mode-2 droplets, and a combined Rayleigh/mode-1/mode-2 particle.
      \textbf{C}. Phase curve for $\alpha$$\ge$130$^{\circ}$. 
      Exploration of the model sensitivity to $H$ (=2, purple; 4, cyan) and 
      $\Delta$ (=4.7$H$, dashed; $\infty$, dashed-dotted).
      \textbf{D}. Phase curve for $\alpha$$\le$50$^{\circ}$. 
}
   \end{figure*}

For our analysis, we adopted published 
phase curves for disk-integrated photometry of Venus in four band filters 
(B, V, R, I) (Mallama et al., \cite{mallamaetal2006}). 
The phase curves largely rely on Earth-based observations, 
but for the B and V bands they also include measurements from 
the space solar observatory SOHO 
when the planet is near superior or inferior conjunctions. 
A similarly obtained phase curve in V band is available for Mercury 
(Mallama et al., \cite{mallamaetal2002}), 
which has become a standard for investigating Mercury's opposition surge
(Domingue et al., \cite{domingueetal2010}).
The measured Venus phase curve for the B band
is presented in Fig. \ref{LCmallama_fig}-A (symbols) in the normalised form 
$A_g$$\Phi$($\alpha$) for the 0--180$^{\circ}$-range of the phase angle $\alpha$. 
$A_g$ is the geometric albedo and measures the planet's fully illuminated  brightness
relative to the brightness of a Lambertian disk of identical cross section. 
The planet phase function, $\Phi$($\alpha$) (with $\Phi$(0)$\equiv$1), 
describes the planet's disk-integrated efficiency
for scattering the incident starlight at varying phases on the orbit.

To aid with the interpretation, Fig. \ref{LCmallama_fig}-A presents 
the phase curve (solid blue line) at the wavelength $\lambda$=0.445 $\mu$m 
as predicted with a multiple-scattering radiative-transport model based on 
backward Monte Carlo integration (see online material). 
For reference, Fig. \ref{LCmallama_fig}-B shows the scattering phase function  
(i.e. the F$_{11}$ element of the scattering matrix) 
for Rayleigh particles, for the so-called mode-1 and mode-2 droplets
that dominate the Venus upper cloud, and for a combined Rayleigh/mode-1/mode-2 particle,  
as obtained with Mie scattering theory at $\lambda$=0.445 $\mu$m. 
Figure \ref{LCmallama_fig}-A generally describes the 
dimming of Venus as its illuminated fraction 
decreases from $\alpha$=0 to 180$^{\circ}$. 
Two features (I and II, marked in the graph)  
stand out, one at each end of the phase curve.

The local brightness enhancement noted as feature I 
is caused by grazing starlight scattered
towards the observer  (Mallama et al., \cite{mallamaetal2006}). 
The forward-scattering peak of mode-2 droplets 
partly compensates for the shrinking size of the illuminated disk
at those phases.
Near inferior conjunction, 
scattered starlight creates a diffuse halo
at altitudes where the atmosphere 
becomes optically thick in limb viewing 
(Garc\'ia Mu\~noz \& Mills, \cite{garciamunozmills2012}). 
The halo eventually encircles the planet and contributes effectively to its brightness 
out to $\alpha$=180$^{\circ}$. 
The brightness strongly depends  on the vertical stratification of the scattering particles
in the atmosphere (see online material and Fig. \ref{LCmallama_fig}-C).
Historically, the varying extension of the halo cusps was key 
to establishing the existence of a Venus atmosphere (Russell, \cite{russell1899}). 
This diffuse halo differs from the refraction halo that 
becomes visible during the ingress and egress phases of a 
Venus transit across the solar disk 
(Garc\'ia Mu\~noz \& Mills, \cite{garciamunozmills2012};
Tanga et al., \cite{tangaetal2012}).
Feature I must not be mistaken for specular reflection at
the planet surface, a mechanism that might also produce brightness excesses at large
phase angles for planets with a liquid surface (Williams \& Gaidos,
\cite{williamsgaidos2008}; Robinson et al., \cite{robinsonetal2010}).

\section{Venus glory}

Feature II provides evidence for a so-called (backward) glory 
(Adam, \cite{adam2002}; van de Hulst, \cite{vandehulst1981}).
It manifests itself in the empirical $A_g$$\Phi$($\alpha$) curve of
Figs. \ref{LCmallama_fig}-A and D as a distinct local
maximum near $\alpha$$\sim$11$^{\circ}$ and a drop in the planet
brightness towards smaller phase angles. 
The SOHO measurements (limited to $\alpha$=6--11$^{\circ}$, 
separate colour code) confirm that trend. 
The modelled phase curve describes the glory in more detail 
as the combination of a local minimum at $\alpha_{\rm{m}}$$\sim$6$^{\circ}$
and a local maximum at $\alpha_{\rm{M}}$$\sim$11$^{\circ}$. 
The theoretical contrast $\Phi(\alpha_{\rm{M}})$/$\Phi(\alpha_{\rm{m}})$ between 
the two local features is $\sim$1.10.
This is the first instance that a glory is identified
in the disk-integrated photometry of a planet.
 
In atmospheric optics, 
a glory is a phenomenon associated with 
light rays that interfere within spherical droplets 
through both internal reflection and surface waves that 
emerge subsequently close to the backscattering direction 
(Adam, \cite{adam2002}; van de Hulst, \cite{vandehulst1981}). 
On Earth, glories are seen above liquid water clouds (Sassen et al., \cite{sassenetal1998}). 
In the Venus atmosphere, 
localised glories at wavelengths from the ultraviolet to the near-infrared 
have recently been spotted from orbit
by the ESA Venus Express mission (Markiewicz et al., \cite{markiewiczetal2014}).
There is a clear resemblance between their reported local reflectance curves and
the disk-integrated feature II of our Figs. \ref{LCmallama_fig}-A and D. 
Glories are explained by Mie scattering theory
and become inhibited by slight departures from sphericity 
of the scattering particles 
(Sassen et al., \cite{sassenetal1998}; 
Mishchenko et al., \cite{mishchenkoetal1996}).
This property makes glories indicators for liquid clouds 
because only liquid condensates are likely to adopt spherical shapes 
in naturally occurring clouds.

   \begin{figure}
   \centering
   \includegraphics[width=9.2cm]{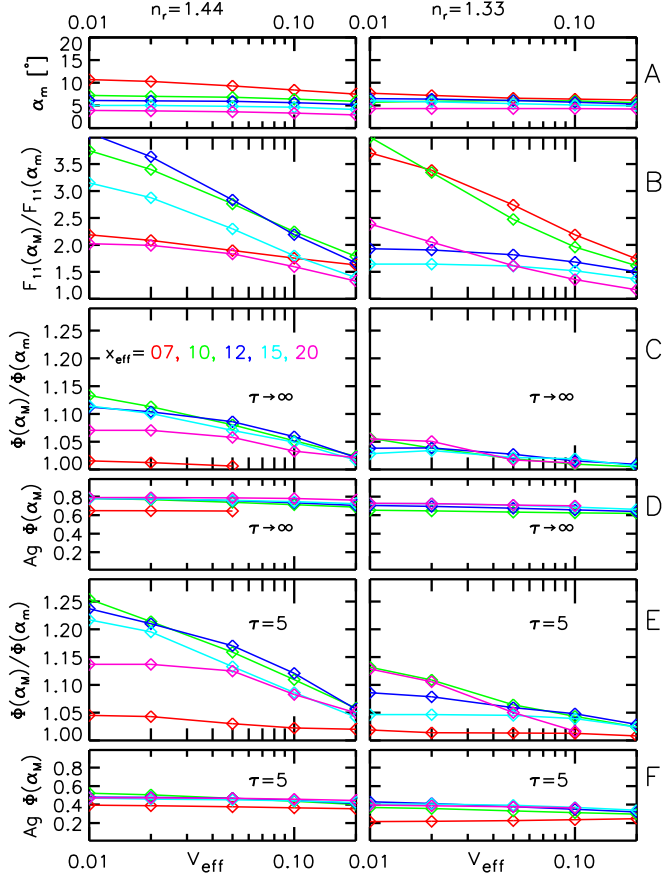}
\caption{\label{contrast_fig}
Investigation of the glory in the space of
$x_{\rm{eff}}$ and $v_{\rm{eff}}$ parameters. 
Left and right panels represent
calculations for refractive indices $n_r$=1.44 and 1.33, respectively. 
Panels \textbf{A}--\textbf{B} are the positions of the local minimum within the glory, 
$\alpha_{\rm{m}}$, 
and F$_{11}$($\alpha_{\rm{M}}$)/F$_{11}$($\alpha_{\rm{m}}$) is the associated 
contrast based on the scattering phase function. 
Panels \textbf{C}--\textbf{F} incorporate the multiple scattering calculations for the prescribed 
Venus-based atmospheres. 
Panels \textbf{C}--\textbf{D} are for semi-infinite atmospheres, 
whereas \textbf{E}--\textbf{F} are for atmospheres of
finite optical thickness equal to 5 above a black surface.
All calculations assume a uniform atmosphere of the specified unimodal size
distribution of scattering droplets. In Panels \textbf{C}--\textbf{F}, 
only configurations leading to $\Phi(\alpha_{\rm{M}})$/$\Phi(\alpha_{\rm{m}})$$>$1 are shown. 
At $\lambda$=0.445 $\mu$m, $x_{\rm{eff}}$ for Venus and Earth 
(average effective radius $\sim$6 $\mu$m) liquid clouds are $\sim$14 and 85, respectively. 
}
\end{figure}

Taking the Venus clouds as a reference, we explore in Fig. \ref{contrast_fig}
the magnitudes for $\alpha_{\rm{m}}$, 
F$_{11}$($\alpha_{\rm{M}}$)/F$_{11}$($\alpha_{\rm{m}}$), 
$\Phi(\alpha_{\rm{M}})$/$\Phi(\alpha_{\rm{m}})$ and 
$A_g$$\Phi$($\alpha_{\rm{M}}$) in the space of 
$x_{\rm{eff}}$(=2$\pi$$r_{\rm{eff}}$/$\lambda$) and $v_{\rm{eff}}$ parameters
that characterise the droplet size distributions 
(see online material for a description of these parameters). 
When multiple local maximae and minimae are present  
(typically for the smaller $v_{\rm{eff}}$ and 
$x_{\rm{eff}}$$\ge$15), 
$\alpha_{\rm{m}}$ and $\alpha_{\rm{M}}$ specify the phase angles 
closest to backscattering.
The ratio F$_{11}$($\alpha_{\rm{M}}$)/F$_{11}$($\alpha_{\rm{m}}$) 
is included because it modulates the contrast. 
Left and right panels correspond to refractive indices 
$n_{\rm{r}}$=1.44 and 1.33, respectively, 
the latter being broadly representative of water at visible wavelengths. 
Liquid water is a pre-requisite for life, 
and liquid water clouds might suggest surface water on a planet
(Bailey, \cite{bailey2007}).

Figs. \ref{contrast_fig}-A show that
the glory concentrates at a location close to backscattering as the 
droplet effective radius increases. 
Theory predicts that the glory angular width 
is proportional to 1/$x_{\rm{eff}}$ (Adam, \cite{adam2002}). 
As a rule (Figs. \ref{contrast_fig}-B), 
F$_{11}$($\alpha_{\rm{M}}$)/F$_{11}$($\alpha_{\rm{m}}$)
increases as $v_{\rm{eff}}$ decreases,  
meaning that narrow size distributions lead to stronger 
contrasts. 
The dependence of F$_{11}$($\alpha_{\rm{M}}$)/F$_{11}$($\alpha_{\rm{m}}$) 
on $x_{\rm{eff}}$ is less obvious. 
For either sufficiently large or sufficiently small particles 
the glory becomes smeared out, 
and there appears to be an optimum $x_{\rm{eff}}$ ($\sim$10--15 for $n_r$=1.44) 
that maximizes 
F$_{11}$($\alpha_{\rm{M}}$)/F$_{11}$($\alpha_{\rm{m}}$). 
For Venus, $x_{\rm{eff}}$$\sim$14.1 at $\lambda$=0.445 $\mu$m, which is
incidentally within the quoted range for the highest glory contrasts.

Figs. \ref{contrast_fig}-C to F investigate the effect of multiple scattering 
on the planet phase curves, which is to attenuate the 
contrast in the disk-integrated glory. 
For a Venus-thick atmosphere (effectively, $\tau$$\rightarrow$$\infty$) the contrast 
$\Phi(\alpha_{\rm{M}})$/$\Phi(\alpha_{\rm{m}})$ 
drops to 1.15 or lower from the higher contrasts suggested by 
F$_{11}$($\alpha_{\rm{M}}$)/F$_{11}$($\alpha_{\rm{m}}$) 
and sometimes the glory pattern becomes undistinguishable. 
The model predicts stronger contrasts in thinner atmospheres that lie above a dark
surface  (compare Figs. \ref{contrast_fig}-C and E, for $\tau$=5), 
at the expense of the planet appearing darker (compare 
Figs. \ref{contrast_fig}-D and F).
Changes in the glory contrast with $\tau$ would be different in 
polarisation because the polarised signal is especially sensitive to the top of the atmosphere 
(Hansen \& Hovenier, \cite{hansenhovenier1974}; Markiewicz et al., \cite{markiewiczetal2014}).
The refractive index has a distinct impact on the glory properties.

\begin{figure}
\centering
\includegraphics[width=8.5cm]{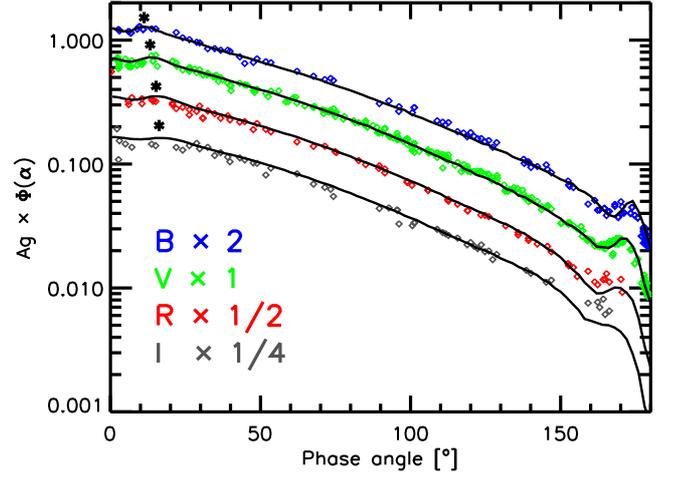}
\caption{\label{bvri_fig}
Venus phase curves in the B, V, R and I bands. 
The glory shifts towards larger phase angles at longer wavelengths. A forward-scattering peak
is clearly discernible in the B, V and R bands. Stars mark the shift of the glory towards larger 
phase angles for longer wavelengths.
}
\end{figure}

We extended the phase curve analysis to all bands in Fig. \ref{bvri_fig}.  
The model satisfactorily reproduces the V, R and I phase curves 
without further adjustments in the droplet size distributions and 
predicts that a distinct forward-scattering peak occurs in V and R as well. 
The measurements confirm the peak in V but do not allow for a clear conclusion in R.
The glory pattern is unequivocally discerned 
in the empirical phase curves for the B and V bands, 
and appears as a local change in the curve trends for the R and I bands. 
The angular width of the glory shifts towards larger phase angles 
as $x_{\rm{eff}}$ decreases or, for fixed droplet sizes, as $\lambda$ increases.
The calculated geometric and spherical albedos for each modelled phase
curve and the corresponding 
$A_g$$\Phi(\alpha_{\rm{m}})$ and $\Phi(\alpha_{\rm{M}})$/$\Phi(\alpha_{\rm{m}})$ 
are listed in Table \ref{albedos_table}. 
We note the potential limitations inherent to estimating the geometric 
and spherical
albedos from measured phase curves with an incomplete coverage of the smaller
phase angles.

\begin{table}
\caption{Photometric properties calculated for the modelled phase curves at all four
filters. }             
\label{albedos_table}      
\centering                          
\begin{tabular}{c c c c c c c}        
\hline                
Band & $\lambda$ & $\alpha_{\rm{m}}$ & $A_g$$\Phi(\alpha_{\rm{m}})$ & $\Phi(\alpha_{\rm{M}})$ & \multicolumn{2}{c}{Albedo}  \\    
 &  [$\mu$m] & [$^{\circ}$]  
&  & /$\Phi(\alpha_{\rm{m}})$
& $A_g$ & Spherical \\    

\hline                        
   B & 0.445 &5.7  & 0.58 & 1.10 & 0.63 & 0.79  \\      
   V & 0.550 &6.6  & 0.67 & 1.08 & 0.71 & 0.92  \\
   R & 0.658 &7.0  & 0.66 & 1.07 & 0.70 & 0.93  \\
   I & 0.806 &10.3 & 0.63 & 1.02 & 0.66 & 0.93  \\
\hline                                   
\end{tabular}
\end{table}

\section{Glories in the solar system. Venus as an exoplanet}

Earth hosts liquid water clouds, and Titan very likely has liquid methane
clouds that lead to occasional rain 
(Hueso \& S\'anchez-Lavega, \cite{huesosanchezlavega2006};
Tokano et al., \cite{tokanoetal2006}).
Phase curves of Earth covering phases near full illumination have not  
been obtained (Mallama, \cite{mallama2009}), but high-lying
ice clouds (Karalidi et al., \cite{karalidietal2012}) 
and photometric variability due to weather patterns 
(Karalidi et al., \cite{karalidietal2012}; 
Livengood et al., \cite{livengoodetal2011}) will probably mask a virtual glory.
For Titan, the optically thick haze and ice condensates above 
the sparse, tropospheric liquid methane clouds 
will almost certainly prevent potential glories 
from being viewed from afar. 
On Jupiter, liquid water clouds are predicted at pressure levels of a
few bars (Carlson et al., \cite{carlsonetal1988}; Zuchowski et al., \cite{zuchowskietal2009}),
but the overlaying condensates will certainly remove the glory feature
(Stam et al., \cite{stametal2004}).

Clouds and haze are present on exoplanets 
over a broad range of planet sizes and orbital distances 
(Currie et al., \cite{currieetal2011};
Demory et al., \cite{demoryetal2013};
Evans et al., \cite{evansetal2013};
Knutson et al., \cite{knutsonetal2014};
Kreidberg et al., \cite{kreidbergetal2014}) 
and are becoming major hurdles in the investigation of 
exoplanet atmospheres. 
It is thus relevant to assess the detectability of exoplanet glories 
because they might potentially indicate extant liquid clouds. 
The brightness contrast between a planet and its host star is
$F_p$($\alpha$)/$F_{\star}$=($R_p$/$a_{\star}$)$^2$$A_g$$\Phi(\alpha)$, where
$R_p$ and $a_{\star}$ are the planet radius and the planet-star distance
respectively. 
For Venus in V band, $A_g$$\Phi(\alpha_{\rm{m\;(,M)}})$=0.67(, 0.73), which leads 
to $F_p$($\alpha_{\rm{m\;(,M)}}$)/$F_{\star}$=2.05(, 2.2)$\times$10$^{-9}$. 
The difference, $\sim$1.5$\times$10$^{-10}$, is beyond what can currently be 
attempted with unresolved photometry of the planet-star system or with ground-based direct
imaging, 
but is within the capacities targeted by the envisioned Terrestrial Planet Finder (TPF)
space missions that may some day directly image exoplanets
in reflected starlight 
(Cash, \cite{cash2006}; Spergel, \cite{spergeletal2013}; 
Traub \& Oppenheimer, \cite{trauboppenheimer2010}). 
Direct imaging however is limited to angular separations 
subtended by the planet-star system, $\theta$,  
larger than the instrument inner working angle, IWA.
Assuming that the planet-star system is seen edge on and at a convenient distance of 5 pc, 
$\theta$($\alpha_{\rm{m}}$)$\sim$18$\times$10$^{-3}$ arcsec in V band. 
This is smaller than the IWAs$\sim$50--75$\times$10$^{-3}$ arcsec 
contemplated by the proposed TPF missions using 
either internal coronagraphs (Traub \& Oppenheimer, \cite{trauboppenheimer2010}) or 
external occulters (Cash, \cite{cash2006}; Turnbull et al.,
\cite{turnbulletal2012}).
Note that observing at longer wavelengths extends both 
$\alpha_{\rm{m}}$ and the IWA (proportional to $\lambda$ for the
internal coronagraph configuration) and thus cancels 
a potential gain in an increased $\alpha_{\rm{m}}$. 
The IWA is largely dictated by the occulter-to-telescope distance
for the external occulter configuration.
The online material discusses the prospects for glory detection on exoplanets in more detail.

We reported the first ever evidence for a glory in the disk-integrated
photometry of Venus and, in turn, of any planet. 
Because glories arise from scattering within spherical droplets, 
they are valuable proxies for extant liquid clouds. 
The detection of exoplanet glories will  admittedly be challenging and calls for a 
TPF space mission.  
The potential pay-off, however, is remarkable because a glory 
detection will yield 
insight into the droplet size distribution and the cloud 
microphysics and, therefore, into the chemical, 
thermal, and dynamical conditions in the planet atmosphere. 
High-precision, multi-wavelength measurements of $\alpha_{\rm{m}}$
and $\Phi(\alpha_{\rm{M}})$/$\Phi(\alpha_{\rm{m}})$
might in principle give access to the main four 
cloud/atmospheric parameters 
that shape the phase curve near superior conjunction, 
$r_{\rm{eff}}$, $v_{\rm{eff}}$, $n_{\rm{r}}$, and $\tau$,
in a way similar to the retrieval of cloud
properties in Earth glory observations (Mayer et al., \cite{mayeretal2004}). 
However, the technical requirements for capturing such subtleties are 
beyond the reach of projected exoplanet characterisation missions for the near future.
Finally, it is fair to state that the lessons learned from Venus
may prove most valuable in the investigation of exoplanet clouds. 

\begin{acknowledgements}

We gratefully acknowledge critical readings of the manuscript by Kate G. Isaak, 
Ana Heras and Dima V. Titov 
(all at ESA/ESTEC, the Netherlands) and the constructive comments from Nick Cowan.

\end{acknowledgements}

\newpage

\clearpage

\appendix

\section{Methods}

\begin{figure*}
   \centering
   \includegraphics[width=9.cm]{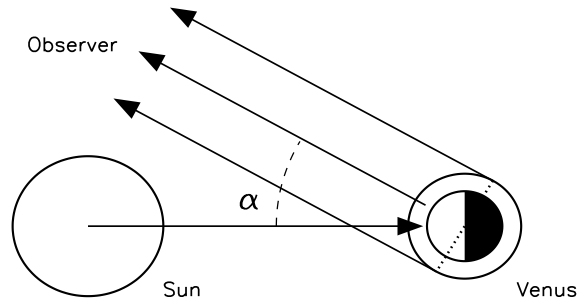}
   \caption{\label{disk_fig} Integration geometry for the backward Monte Carlo
   algorithm. Photons are traced back from the observer towards the planet. 
   The integration domain encompasses the entire planet disk projected from the
   observer site, whether directly illuminated by the star or not. 
   }
   \end{figure*}

The backward Monte Carlo model used here
(Garc\'ia Mu\~noz \& Pall\'e, \cite{garciamunozpalle2011};
Garc\'ia Mu\~noz et al., \cite{garciamunozetal2011}) 
calculates the starlight reflected from the planet full disk
as a function of the star-planet-observer phase angle (Fig. \ref{disk_fig}),  
including the crescent that forms at the edges of the disk approaching inferior conjunction.
The model computes the four elements of the Stokes vector 
and properly handles the planet curvature.

\begin{figure*}
   \centering
   \includegraphics[width=9.cm]{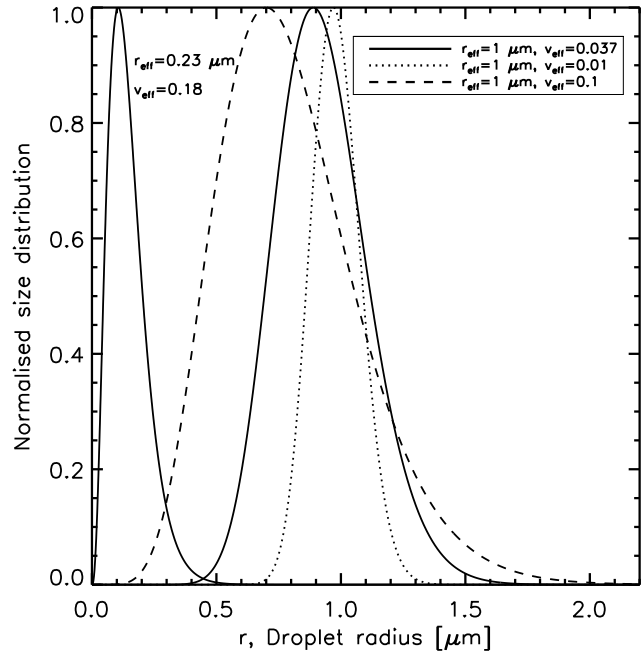}
   \caption{\label{gamma_fig} Normalised gamma distribution for radius sizes of
   mode-1 and mode-2 droplets (solid curves). 
   The glory is very sensitive to the effective variance, 
   $v_{\rm{eff}}$, which is a measure of the size distribution width.  
   }
   \end{figure*}

In the calculations, we imposed that the Venus atmosphere is effectively semi-infinite
(optical thickness, $\tau$$\rightarrow$$\infty$) 
and cloud extinction is contributed in a 0.2:1 ratio by
a bimodal admixture of so-called mode-1
(effective radius $r_{\rm{eff}}$=0.23 $\mu$m, 
and effective variance $v_{\rm{eff}}$=0.18), 
and mode-2 ($r_{\rm{eff}}$=1 $\mu$m, $v_{\rm{eff}}$=0.037)
spherical droplets of H$_2$SO$_4$/H$_2$O (75/25\% by mass)
with real refractive indices $n_{\rm{r}}$=1.45 (B), 1.44 (V, R) and 1.43 (I). 
We also included Rayleigh scattering in a proportion
0.045$\times$(0.55/$\lambda$[$\mu$m])$^4$ relative to cloud extinction. 
The droplet properties were calculated from Mie theory (Mishchenko et
al., \cite{mishchenkoetal2002}) assuming gamma distributions  
for the particle sizes
(particle radius density $\propto$$r^{(1-3v_{\rm{eff}})/v_{\rm{eff}}}$$\exp{(-r/(r_{\rm{eff}}v_{\rm{eff}}))}$). 
The dimensionless parameters 
$x_{\rm{eff}}$(=2$\pi$$r_{\rm{eff}}$/$\lambda$) and $v_{\rm{eff}}$ are
particularly useful to prescribe the droplet optical properties, 
including the $F_{11}$ element of the scattering matrix
(Hansen \& Travis, \cite{hansentravis1974}). 
Figure \ref{gamma_fig} shows the normalised distributions of particle sizes 
for both mode-1 and mode-2 droplets.

\begin{figure*}
   \centering
   \includegraphics[width=9.cm]{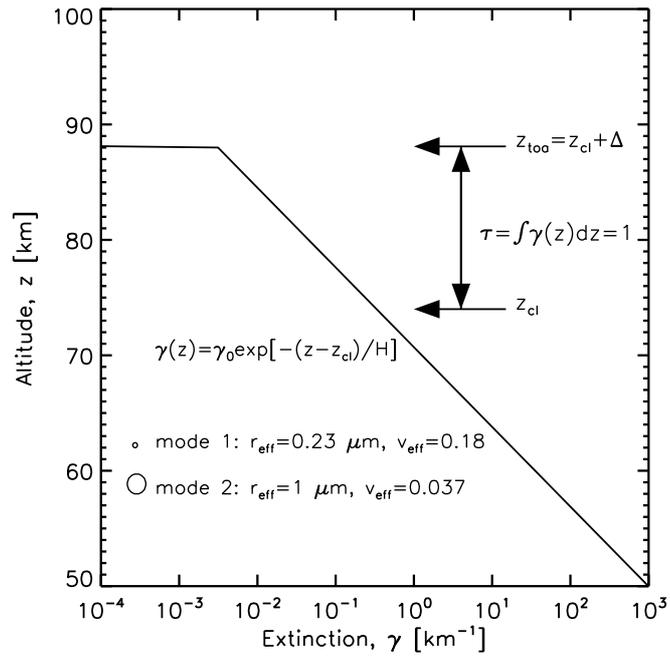}
   \caption{\label{tausketch_fig}
   Sketch that illustrates the meaning of the parameters in our prescribed model
   atmosphere.}
   \end{figure*}

Total extinction decays exponentially in the vertical with a scale height 
$H$ up to a level $z_{\rm{toa}}$=$z_{\rm{cl}}$+$\Delta$ above the level
$z_{\rm{cl}}$ ($\sim$64--74 km in the Venus atmosphere) of nadir-integrated optical depth 
equal to one, and is taken to be zero upwards from there. 
$\Delta$ effectively truncates the cloud top without the need to invoke a more elaborate
multi-layer description of the Venus upper cloud and haze  (Fig.
\ref{tausketch_fig}). 
Photons were assumed to scatter without absorption at the wavelengths of the V (0.55 $\mu$m), 
R (0.658 $\mu$m) and I (0.806 $\mu$m) filters.
In B band (0.445 $\mu$m), however, we adopted a uniform single-scattering albedo of 0.9975 that shifts 
the modelled phase curve down with respect to the fully conservative configuration. 
This modification accounts for the unknown absorber that 
confers  the patchy appearance seen in imagery at $\lambda$$\le$0.5 $\mu$m on the planet
(Titov et al., \cite{titovetal2008}).
Our prescriptions provide a simple but realistic 
description for the upper-cloud region of the Venus atmosphere,
with parameter values traceable to various references 
(Hansen \& Hovenier, \cite{hansenhovenier1974}; Crisp, \cite{crisp1986},
Pollack et al., \cite{pollacketal1980}), and is overall consistent with the 
prescriptions used in recent 
investigations of the Venus upper cloud (Ignatiev et al., \cite{ignatievetal2009};
 Wilquet et al., \cite{wilquetetal2012}). 

Numerical experiments 
show that for $\alpha$$\le$150$^{\circ}$ the planet brightness is mainly dictated by the total 
optical thickness of the atmosphere and the optical properties of the scattering particles. 
The good match between the empirical and modelled phase curves 
(after adjusting the atmospheric single-scattering albedo at $\lambda$=0.445 $\mu$m) 
for our standard description of combined mode-1 and mode-2 droplets 
does not warrant a full exploration of the specific droplet parameters. 
Thus, the calculations presented in Figs. \ref{LCmallama_fig} and \ref{bvri_fig} make
unmodified use of these prescriptions.
For $\alpha$$\ge$150$^{\circ}$, however, the brightness becomes very sensitive 
to the vertical distribution of the scattering particles and thus to $H$ and $\Delta$. 
We explored $H$ and $\Delta$, reaching the solutions of Figs. \ref{LCmallama_fig}-A
and \ref{bvri_fig} for $H$=3 km and $\Delta$=4.7$H$=14.1 km from the visual adjustment of
the measured phase curves in the B and V bands, with priority given to SOHO data. 
Again, these choices are overall consistent with recent investigations of the Venus upper cloud 
(Ignatiev et al., \cite{ignatievetal2009}; Wilquet et al., \cite{wilquetetal2012}). 
Furthermore, Fig. \ref{LCmallama_fig}-C demonstrates the sensitivity of the 
forward-scattering peak to
the two parameters with predicted phase curves for selected values of $H$ and $\Delta$. 
As a final check, we compared our modelled linear polarisations with 
available disk-integrated measurements (Hansen \& Hovenier, \cite{hansenhovenier1974}); 
the result of the comparison (not shown) confirmed the validity of the prescribed atmosphere.

\section{Exoplanetary glories?}

On a more speculative basis, one may devise 
more favourable conditions for detecting exoplanet glories, 
at least theoretically. 
As a rule, larger planetary radii result in brighter planets that are
more likely to stand above the stellar glare. 
For direct imaging, 
longer orbital distances would lead to larger angular separations, e.g. 
$\theta$($\alpha_{\rm{m}}$)$\sim$50$\times$10$^{-3}$ arcsec for $a_{\star}$=2 AU. 
Planets at such distances might conceivably allow for liquid- 
(rather than solid-) particle clouds if, 
for instance, the host star irradiates strongly and thus keeps the planet warm, 
if the planet is still contracting and therefore heated from below or 
if it has been externally heated by a catastrophic event 
(Lupu et al., \cite{lupuetal2014}).
In combined planet-star light, in contrast, 
conditions suitable for a glory detection seem better 
for close-in planets around faint stars, 
which might ensure warm temperatures and acceptable contrasts
at small $a_{\star}$ distances.
Planets orbiting white dwarfs represent an example of the latter 
as they might host liquid water at the surface 
(and, arguably, also in the atmosphere) for orbital distances as small as
0.005 AU (Agol, \cite{agol2011}). 
Scaling the Venus glory contrast leads to
$\sim$1.5$\times$10$^{-10}$$\times$(0.73/0.005)$^2$$\sim$3.2$\times$10$^{-6}$,
which is more stringent but not far from precisions of 10$^{-5}$ obtained by CoRoT and Kepler. 
For a super-Venus planet twice (thrice) as large as Venus, 
the contrast increases to 1.3(2.9)$\times$10$^{-5}$. 
Photon collection and instrumental stability will be critical aspects of such 
endeavours. 
Obviously, 
these considerations overlook whether Venus-like glory-forming clouds 
might form in the conditions we described 
(or in other planet-star systems that might be postulated), 
a question that is difficult to address because major
uncertainties remain in the processes that govern cloud formation
(Cahoy et al., \cite{cahoyetal2010}; 
Gao et al., \cite{gaoetal2014}; 
Zsom et al., \cite{zsometal2012}).
Glories occur near full illumination, which effectively averages the 
cloud properties over the planet disk. 
This characteristic will likely prevent false positives caused by non-uniformities
in the planet reflective properties (Cowan et al., \cite{cowanetal2013}).

\newpage

\begin{thebibliography}{}

\bibitem[2002]{adam2002} 
Adam, J.A. 2002,
Phys. Rep., 356, 229.

\bibitem[1968]{arkingpotter1968} 
Arking, A. \& Potter, J. 1968, 
J. Atmos. Sci., 25, 617.

\bibitem[2007]{bailey2007} 
Bailey, J. 2007, 
Astrobiol., 7, 320.

\bibitem[1988]{carlsonetal1988} 
Carlson, B.E.,
Rossow, W.B. \&
Orton, G.S. 1988,
J. Atmos. Sci., 45, 2066.

\bibitem[2006]{cash2006}
Cash, W. 2006,
Nature, 442, 51.

\bibitem[1979]{coffeen1979}
Coffeen, D.L. 1979, 
J. Opt. Soc. Am., 69, 1051.

\bibitem[2011]{currieetal2011} 
Currie, T.,
Burrows, A.,
Itoh, Y.,
Matsumura, S.,
Fukagawa, M., et al. 2011, 
Astrophys. J., 729, 128.

\bibitem[2013]{demoryetal2013} 
Demory, B.-O.,
de Wit, J.,
Lewis, N.,
Fortney, J.,
Zsom, A., et al. 2013,
Astrophys. J. Lett., 776, L25.

\bibitem[2010]{domingueetal2010}
Domingue, D.L.,
Vilas, F.,
Holsclaw, G.M.,
Warell, J.,
Izenberg, N.R. et al. 2010, 
Icarus, 209, 101.

\bibitem[1983]{espositoetal1983}
Esposito, L.W., 
Knollenberg, R.G., 
Marov, M.Y., 
Toon, O.B., \& 
Turco, R. P. 1983, 
The clouds are hazes of Venus. 
In Venus, Eds. D.M. Hunten, L. Colin, T.M. Donahue, \& V.I. Moroz
(Tucson: Univ. of Arizona Press), 484.

\bibitem[2013]{evansetal2013} 
Evans, T.M, 
Pont, F.,
Sing, D.K.,
Aigrain, S.,
Barstow, J.K., et al. 2013,
Astrophys. J. Lett., 772, L16.

\bibitem[2012]{garciamunozmills2012} 
Garc\'ia Mu\~noz, A. \&
Mills, F.P. 2012, 
A\&A, 547, A22. 

\bibitem[1971]{hansenarking1971} 
Hansen, J.E. \&
Arking, A. 1971, 
Science, 171, 669.

\bibitem[1974]{hansenhovenier1974} 
Hansen, J.E. \&
Hovenier, J.W. 1974, 
J. Atmos. Sci., 31, 1137.

\bibitem[1974]{hansentravis1974} 
Hansen, J.E. \&
Travis, L.D. 1974, 
Space Sci. Rev., 16, 527.

\bibitem[2006]{huesosanchezlavega2006}
Hueso, R. \&
S\'anchez-Lavega 2006,
Nature, 442, 428.

\bibitem[2012]{karalidietal2012} 
Karalidi, T.,
Stam, D.M. \& Hovenier, J.W. 2012,
A\&A, 548, A90. 

\bibitem[1977]{knollenbergetal1977} 
Knollenberg, R.G.,
Hansen, J.,
Ragent, B.,
Martonchik, J. \&
Tomasko, M. 1977, 
Space Science Rev., 20, 329.

\bibitem[2014]{knutsonetal2014} 
Knutson, H.A.,
Benneke, B.,
Deming, D. \&
Homeier, D. 2014, 
Nature, 505, 66.

\bibitem[2014]{kreidbergetal2014} 
Kreidberg, L.,
Bean, J.L.,
D\'esert, J.-M.,
Benneke, B.,
Deming, D. et al. 2014, 
Nature, 505, 69.

\bibitem[2011]{livengoodetal2011} 
Livengood, T.A.,
Deming, L.D.,
A'Hearn, M.F.,
Charbonneau, D.,
Hewagama, T. et al. 2011, 
Astrobiol., 11, 907.

\bibitem[2002]{mallamaetal2002} 
Mallama, A.,
Wang, D. \&
Howard, R.A. 2002, 
Icarus, 155, 253.

\bibitem[2006]{mallamaetal2006} 
Mallama, A.,
Wang, D. \&
Howard, R.A. 2006, 
Icarus, 182, 10.

\bibitem[2009]{mallama2009} 
Mallama, A. 2009, 
Icarus, 204, 11.

\bibitem[2014]{markiewiczetal2014} 
  Markiewicz, W.J.,
  Petrova, E.,
  Shalygina, O.,
  Almeida, M.,
  Titov, D.V. et al. 2014,
  Icarus, 234, 200.

\bibitem[2004]{mayeretal2004} 
Mayer, B., 
Schr\"oder, M., 
Preusker, R. \&
Sch\"uller, L. 2004,
Atmosph. Chem. and Phys., 4, 1255

\bibitem[1996]{mishchenkoetal1996} 
Mishchenko, M.I.,
Travis, L.D. \&
Macke, A. 1996,
Appl. Optics, 35, 4927.

\bibitem[2010]{robinsonetal2010} 
Robinson, T.D.,
Meadows, V.S. \&
Crisp, D. 2010,
Astrophys. J. Lett., 721, L67.

\bibitem[1978]{rossow1978}
Rossow, W.B. 1978, 
Icarus, 36, 1.

\bibitem[1899]{russell1899} 
Russell, H.N. 1899, 
Astrophys. J., 9, 284.

\bibitem[2004]{sanchezlavegaetal2004} 
S\'anchez-Lavega, A.,
P\'erez-Hoyos, S. \&
Hueso, R. 2004,
Am. J. Phys., 72, 767.

\bibitem[1998]{sassenetal1998} 
Sassen, K.,
Arnott, W.P.,
Barnett, J.M. \&
Aulenbach, S. 1998
Appl. Optics, 37, 1427.

\bibitem[2013]{spergeletal2013} 
Spergel, D.,
Gehrels, N.,
Breckinridge, J.,
Donahue, M.,
Dressler, A., et al. 2013,
Wide-Field InfraRed Survey Telescope-Astrophysics Focused
Telescope Assets WFIRST-AFTA. Final Report. Science definition.

\bibitem[2004]{stametal2004} 
Stam, D.M.,
Hovenier, J.W. \&
Waters, L.B.F.M. 2004,
A\&A, 428, 663. 

\bibitem[2012]{tangaetal2012} 
Tanga, P.,
Widemann, T.,
Sicardy, B.,
Pasachoff, J.M.,
Arnaud, J., et al. 2012,
Icarus, 218, 207.

\bibitem[2006]{tokanoetal2006} 
Tokano, T.,
McKay, C.P.,
Neubauer, F.M.,
Atreya, S.K.,
Ferri, F., et al. 2006,
Nature, 442, 432.

\bibitem[2010]{trauboppenheimer2010} 
Traub, W.A. \& Oppenheimer, B.R. 2010,
Direct imaging techniques. 
In Exoplanets, 
Ed. S. Seager
(University of Arizona Press), 111.

\bibitem[2012]{turnbulletal2012} 
Turnbull, M.C.,
Glassman, T.,
Roberge, A., 
Cash, W., 
Noecker, C.,
Lo, A. et al. 2012, 
PASP, 124, 418.

\bibitem[1981]{vandehulst1981} van de Hulst, H.C. 1981,
Light scattering by small particles
(Dover, New York). 

\bibitem[2008]{williamsgaidos2008} 
Williams, D.M. \&
Gaidos, E. 2008,
Icarus, 195, 927

\bibitem[2009]{zuchowskietal2009} 
Zuchowski, L.C.,
Yamazaki, Y.H. \&
Read, P.L. 2009,
Icarus, 200, 563.

\bibitem[2010]{zuggeretal2010} 
Zugger, M.E., 
Kasting, J.F., 
Williams, D.M., 
Kane, T.J. \&
Philbrick, C.R. 2010,
Astrophys. J., 723, 1168

\bibitem[2011]{zuggeretal2011} 
Zugger, M.E., 
Kasting, J.F., 
Williams, D.M., 
Kane, T.J. \&
Philbrick, C.R. 2011,
Astrophys. J., 739, 12

\end{thebibliography}

\begin{thebibliography}{}

\bibitem[2011]{agol2011} 
Agol, E. 2011,
Astrophys. J., 731, L31.

\bibitem[2010]{cahoyetal2010} 
Cahoy, K.L.,
Marley, M.S. \&
Fortney, J.J. 2010, 
Astrophys. J., 724, 189.

\bibitem[2013]{cowanetal2013}
Cowan, N.B.,
Fuentes, P.A. \&
Haggard, H.M. 2013,
MNRAS, 434, 2465

\bibitem[1986]{crisp1986} 
Crisp, D. 1986,
Icarus, 67, 484.

\bibitem[2014]{gaoetal2014} 
Gao, P.,
Zhang, X., 
Crisp, D.,
Bardeen, C.G. \&
Yung, Y.L. 2014,
Icarus (\textit{in press}).

\bibitem[2011]{garciamunozpalle2011} 
Garc\'ia Mu\~noz, A. \&
Pall\'e, E. 2011, 
JQSRT, 112, 1609.

\bibitem[2011]{garciamunozetal2011} 
Garc\'ia Mu\~noz, A.,
Pall\'e, E.,
Zapatero Osorio, M.R. \&
Mart\'in, E.L. 2011, 
Geophys. Res. Lett., 38, L14805.

\bibitem[2009]{ignatievetal2009} 
Ignatiev, N.I., 
Titov, D. V., 
Piccioni, G.,
Drossart, P., 
Markiewicz et al. 2009, 
J. Geophys. Res., 114, E00B43, doi:10.1029/2008JE003320. 

\bibitem[2014]{lupuetal2014} 
Lupu, R.E., 
Zahnle, K., 
Marley, M.S., 
Schaefer, L., 
Fegley, B. et al. 2014,
arXiv:1401.1499.

\bibitem[2002]{mishchenkoetal2002} 
Mishchenko, M.I.,
Travis, L.D. \&
Lacis, A.A. 2002,
Scattering, absorption, and emission of Light
by small particles (Cambridge University Press, Cambridge).

\bibitem[1980]{pollacketal1980}
Pollack, J.B., 
Toon, O.B., 
Whitten, R.C., 
Boese, R., 
Ragent, B, et al. 1980, 
Distribution and source of the UV absorption in Venus' atmosphere.
J. Geophys. Res., 85, 8141.

\bibitem[2008]{titovetal2008} 
Titov, D.V.,
Taylor, F.W.,
H\aa{}kan, S.,
Ignatiev, N.I.,
Markiewicz, W.J.,
Piccioni, G. \&
Drossart, P. 2008, 
Nature, 456, 620.

\bibitem[2012]{wilquetetal2012} 
Wilquet, V., 
Drummond, R., 
Mahieux, A., 
Robert, S., 
Vandaele, A.C. \&
Bertaux, J.-L. 2012, 
Icarus, 217, 875.

\bibitem[2012]{zsometal2012} 
Zsom, A.,
Kaltenegger, L. \&
Goldblatt, C. 2012, 
Icarus, 221, 603.

\end{thebibliography}
\end{document}